# Radiation Damage Studies on Titanium Alloys as High Intensity Proton Accelerator Beam Window Materials

Taku ISHIDA[1,2*], Eiichi WAKAI[1,3], Shunsuke MAKIMURA[1,2], Patrick G. HURH[4], Kavin AMMIGAN[4], Andrew M. CASELLA[5], Danny J. EDWARDS[5], David J. SENOR[5], Christopher J. DENSHAM[6], Michael FITTON[6], Joe BENNETT[6], Dohyun KIM[7], Nikolaos SIMOS[7], Marco CALVIANI[8], and Claudio TORREGROSA MARTIN[8], on behalf of the RaDIATE Collaboration

[1]*Japan Proton Accelerator Research Complex (J-PARC), Tokai-mura, 319-1106 Japan*
[2]*High Energy Accelerator Research Organization (KEK), Tsukuba, 305-0801 Japan*
[3]*Japan Atomic Energy Agency (JAEA), Tokai-mura, 319-1106 Japan*
[4]*Fermi National Accelerator Laboratory (Fermilab), Batavia, IL 60510-5011 U.S.A.*
[5]*Pacific Northwest National Laboratory (PNNL), Richland, WA 99352 U.S.A.*
[6]*STFC Rutherford Appleton Laboratory (RAL), Oxfordshire, OX11 0QX, U.K.*
[7]*Brookhaven National Laboratory (BNL), Upton, NY 11973-5000 U.S.A.*
[8]*CERN, 1211 Geneva 23, Switzerland*
*E-mail: taku.ishida@kek.jp

A high-strength dual $\alpha+\beta$ phase titanium alloy Ti-6Al-4V is utilized as a material for beam windows in several accelerator target facilities. However, relatively little is known about how material properties of this alloy are affected by high-intensity proton beam irradiation. With plans to upgrade neutrino facilities at J-PARC and Fermilab to over 1 MW beam power, the radiation damage in the window material will reach a few displacements per atom (dpa) per year, significantly above the ~0.3 dpa level of existing data. The international RaDIATE collaboration, Radiation Damage In Accelerator Target Environments, has conducted a high intensity proton beam irradiation of various target and window material specimens at Brookhaven Linac Isotope Producer (BLIP) facility, including a variety of titanium alloys. Post-Irradiation Examination (PIE) of the specimens in the 1st capsule, irradiated at up to 0.25 dpa, is in progress. Tensile tests in a hot cell at Pacific Northwest National Laboratory (PNNL) exhibited a clear signature of radiation hardening and loss of ductility for Ti-6Al-4V, while Ti-3Al-2.5V, with less $\beta$ phase, exhibited less severe hardening. Microstructural investigations will follow to study the cause of the difference in tensile behavior between these alloys. High-cycle fatigue (HCF) performance is critical to the lifetime estimation of beam windows exposed to a periodic thermal stress from a pulsed proton beam. The first HCF data on irradiated titanium alloys are to be obtained by a conventional bend fatigue test at Fermilab and by an ultrasonic mesoscale fatigue test at Culham Laboratory. Specimens in the 2nd capsule, irradiated at up to ~1 dpa, cover typical titanium alloy grades, including possible radiation-resistant candidates. These systematic studies on the effects of radiation damage of titanium alloy materials are intended to enable us not only to predict realistic lifetimes of current beam windows made of Ti-6Al-4V, but also to extend the lifetime by choosing a more radiation and thermal shock tolerant alloy with a preferable heat treatment, or even by developing new materials.

**KEYWORDS:** Titanium Alloy, Beam Window, Proton Beam, Radiation Damage

## 1. Introduction

In the recent past, major accelerator facilities have been limited in beam power by the survivability of beam intercepting devices, such as production targets, beam windows, collimators, and beam dumps. The radiation damage effect on their constituent materials has been identified as the most cross-cutting challenge facing these facilities, one of which is radiation damage to titanium alloys. A high-strength dual phase titanium alloy, Ti-6Al-4V, is used, or is to be adopted, as a material for the:
- J-PARC neutrino facility: primary beam window[1], target containment window[2]
- J-PARC hadron facility: target chamber window[3]
- Long Baseline Neutrino Facility (LBNF) at Fermilab: target containment window[4]
- Facility for Rare Isotope Beams (FRIB) at MSU: beam dump[5]
- ILC: 14 MW main dump beam window[6]

The reason for the choice of the material is due to its high specific strength, ductility, good fatigue endurance limits, and high thermal shock resistance to pulsed beams, which results from a relatively low modulus and small coefficient of thermal expansion. However, relatively little is known on how these excellent properties as beam intercepting devices are affected by high-intensity proton beam irradiation.

Figure 1 shows J-PARC neutrino facility's 1st primary beam window, after its use in the beam-line during 2010 to 2017. It comprises two 0.3 mm-thick partial hemispheres of Ti-6Al-4V, cooled by helium flowing between them. As given in Table I, it was designed to withstand the passage of $3.3 \times 10^{14}$ protons per pulse (ppp) in a few tens of mm$^2$ cross section, which causes localized energy deposition and a calculated periodic thermal stress wave of a few hundred MPa amplitude. When the facility is upgraded to 1.3 MW beam operation with a ~1 Hz repetition rate for the T2K-II[8] and Hyper-Kamiokande[9] projects, it will accumulate $2.4 \times 10^{21}$ protons on target (pot) with about 8-million pulses per operational year. The radiation damage, expressed in terms of displacements per atom (dpa), *i.e.* the average number of displaced atoms in the crystal lattice, is estimated to be about 2 dpa. Significant irradiation hardening and loss of ductility/embrittlement have been reported at only 0.1 dpa, and no data for greater than 0.3 dpa exists[10].

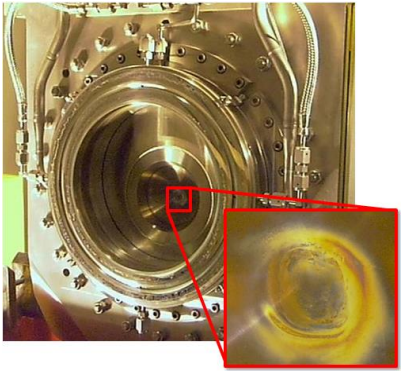

**Fig. 1.** J-PARC neutrino facility's primary beam window, removed from the beam-line in 2017 after experiencing $2.2 \times 10^{21}$ pot[7], similar to the pot/year for 1.3MW operation. Severe discoloration is identified at the beam center.

**Table I.** Parameters of the J-PARC Main Ring upgrade to achieve high power operation by a combination of increasing protons per pulse (ppp) and doubling the repetition rate to about 1 Hz. The ppp necessary for 750kW operation has already been achieved without major problems for the target or the window. Thermal shock is generated by the originally designed $3.3 \times 10^{14}$ ppp, which is similar to the value necessary for 1.3 MW operation at the higher repetition rate. This is the reason why those authors involved consider that the J-PARC neutrino target and window may be suitable for 1.3 MW power with only minor modifications.

| Beam Power | ppp | Cycle | pot/100days |
|---|---|---|---|
| 485kW(achieved) | $2.5 \times 10^{14}$ | 2.48 s | $(0.9 \times 10^{21})$ |
| 750kW(planned) | $2.0 \times 10^{14}$ | 1.3 s | $1.3 \times 10^{21}$ |
| 750kW(designed) | $3.3 \times 10^{14}$ | 2.1 s | $1.3 \times 10^{21}$ |
| 1.3MW(proposed) | $3.2 \times 10^{14}$ | 1.16 s | $2.4 \times 10^{21}$ |

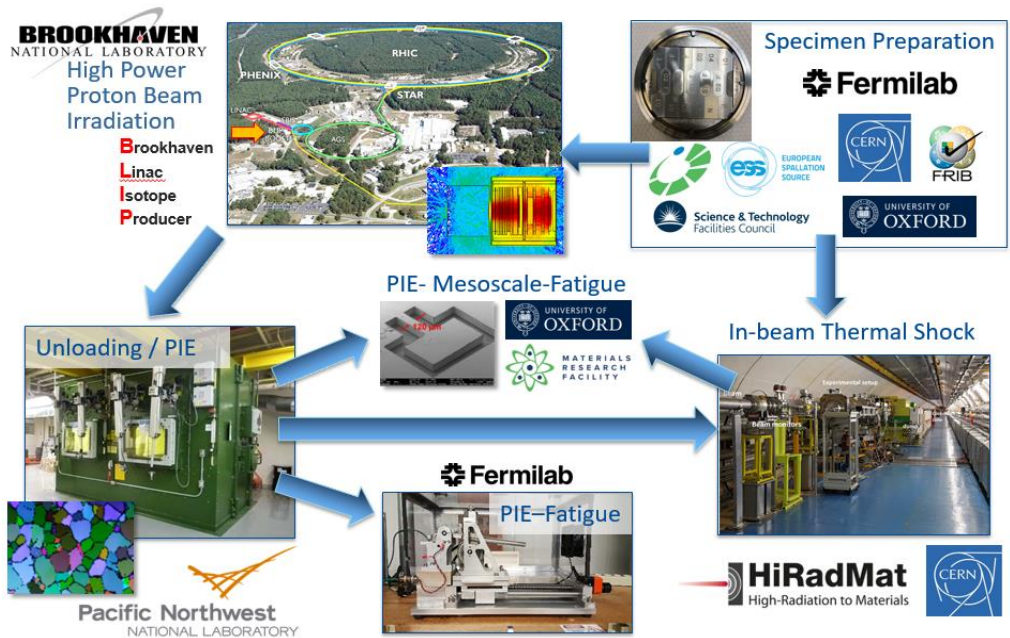

**Fig. 2.** Overview of the RaDIATE research programs.

As described above, the beam window is required to withstand around 10-million thermal stress cycles in its lifetime. However, no known high cycle fatigue (HCF) data exists for this alloy in the irradiation condition more than $10^3$ cycles[11].

The J-PARC neutrino facility employs a 900 mm-long helium gas-cooled isotropic graphite rod as a production target, with Ti-6Al-4V utilized as a containment vessel and a proton beam entrance window[2]. In comparison with water cooling adopted by some other facilities, helium cooling enables the graphite to operate at a reasonably high temperature in order to mitigate radiation damage effects (the temperature should not be too high, since oxidization of graphite caused by impurity in the helium coolant can limit the life of the target). For nominal 750 kW operation, the maximum temperature of graphite is estimated to be 740°C. To maintain the maximum temperature of about 900°C at 1.3MW operation, the helium mass flow will be approximately doubled (32 to 60 g/s) by applying a higher pressure (1.6 bar to 5 bar)[12]. The operational conditions and radiation damage level of the target entrance window will be similar to those for the primary beam window described earlier. Since a similar target concept is to be adopted for the Long Baseline Neutrino Facility (LBNF) at Fermilab[4], the study of radiation damage effects on Ti-6Al-4V, or an alternative grade, is also critical for the successful future operation of that facility.

## 2. High-Power Proton Irradiation at BLIP and Post-Irradiation Examinations

To replicate the severe High Energy Physics (HEP) target environment and provide bulk samples for radiation damage analysis, high energy, high fluence and large volume proton irradiations are required. These runs, including Post-Irradiation Examination (PIE), are expensive and can take a long time. To promote these studies, the RaDIATE international collaboration, Radiation Damage In Accelerator Target Environments[13], was founded in 2012 by 5 institutions led by Fermilab and STFC to bring together the HEP/Basic Energy Science (BES) accelerator target and nuclear fusion/fission materials communities. In 2017, the 2nd MoU revision included J-PARC (KEK, JAEA) and CERN as official participants, and the collaboration has now grown to about 70 members from 14 institutions. The research program consists of determining the effect of high energy proton irradiation on the mechanical and physical properties of potential target and beam window materials, and to understand the underlying changes by advanced microstructural studies. The goal is to enable realistic component lifetime predictions, to design robust multi-MW targets and beam windows, and further, to select the best or even develop new materials to extend their lifetimes. A schematic overview of the programs promoted by the collaboration is given in Fig. 2.

*2.1 High Power Proton Irradiation at BNL BLIP Facility*

One of the major RaDIATE activities is a high-intensity proton irradiation of mechanical/micro-structural test specimens at the Brookhaven Linear Isotope Producer (BLIP) facility, BNL[14]. As shown in Fig. 3, our radiation damage study capsule box is placed upstream of BLIP's medical isotope production target box, where multiple capsules contain different materials (Beryllium, Graphite, Silicon, Aluminum, Titanium and Heavy materials). Capsules are made of stainless steel with thin beam windows, and the outer surfaces of each capsule is cooled by water. Each capsule is filled with numerous specimens (tensile, bend, fatigue, micro-structural studies) sealed in an inert gas or vacuum atmosphere. In total, over 200 specimens in 9 capsules were fabricated by participating accelerator institutions, and provided for the irradiation run at BLIP.

The irradiation campaign was executed in 3 phases during 2017 to 2018 for different configurations with six capsules in each phase. There are three titanium capsules, where an upstream capsule, 'US-Ti' was placed at the $5^{th}$ location for all three irradiation phases. Two downstream capsules, 'DS-Ti1' and 'DS-Ti2', were placed at the most downstream $6^{th}$ location during phase-1 and phase-3, respectively. A rastered beam with ~2.4 cm diameter footprint was delivered to the target box with an average current of 154 µA and a peak fluence of $1.1\times10^{14}$ p/cm$^2$·s. The total energy loss in the RaDIATE target box assembly is limited to 68 MeV out of 181 MeV beam energy, in order to deliver uniform beam energy and flux to the downstream isotope targets.

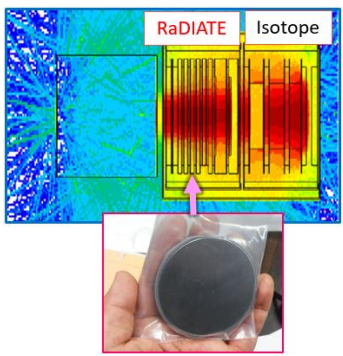

**Fig. 3.** A RaDIATE target box with 6 capsules, locating upstream of a medical isotope production target box.

**Table II.** Summary of the irradiation phases during 2017–2018 on the RaDIATE target box. There are three titanium capsules: DS-Ti1 was irradiated in phase-1, DS-Ti2 in phase-3, and US-Ti during entire 1,2 and 3 phases, respectively. The estimated dpa for each capsule is also given, which is to be compared with the existing data, ~0.3dpa.

|  | 2017 | | 2018 | Phase-1~3 Total |
|---|---|---|---|---|
|  | Phase-1 | Phase-2 | Phase-3 |  |
| Total Hours | 226.3 | 302.9 | 789.1 | 1,318.3 |
| Total Days | 9.43 | 12.62 | 32.88 | 54.93 |
| Av. Curr. (µA) | 143.48 | 150.57 | 158.38 | 154.03 |
| pot ($\times 10^{21}$) | 0.73 | 1.03 | 2.81 | 4.57 |
| Ti capsule | DS-Ti1 | - | DS-Ti2 | US-Ti |
| Max dpa-NRT | 0.25 | - | 0.95 | 1.53 |

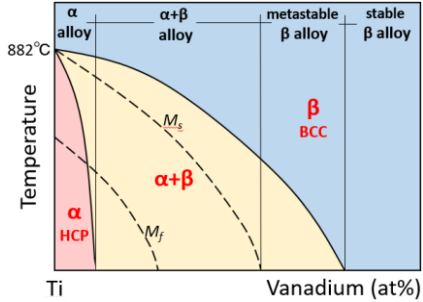

**Fig. 4.** Schematic of the titanium-to-vanadium phase diagram. Vanadium is a typical β stabilizer element for titanium, which lowers the β-transus for pure titanium (882°C). Titanium alloys can be classified into three categories, i.e., α, α+β, and metastable β alloys.

**Table III.** List of titanium alloy grades included in the BLIP irradiation. *Heat Treatment: 'A' stands for mill-annealing, and 'STA' solution treatment and aging.

| ASTM Grade Composition | | HT* | Tensile Properties | | |
|---|---|---|---|---|---|
| | | | Tensile (MPa) | Yield (MPa) | El.. (%) |
| **Commercially Pure (CP) Titanium** | | | | | |
| • Gr-1 | | A | 270~410 | ≧165 | ≧27 |
| • Gr-2 | | A | 340~510 | ≧215 | ≧23 |
| **α alloy** | | | | | |
| • Gr-6 | Ti-5Al-1.5Sn | A | 862 | 804 | 16 |
| **α + β alloy** | | | | | |
| • Gr-9 | Ti-3Al-2.5V | A | 686 | 588 | 20 |
| • Gr-5/Gr-23 ELI | | A | 980 | 921 | 14 |
| | Ti-6Al-4V | STA | 1,170 | 1,100 | 10 |
| **Metastable β alloy** | | | | | |
| • Ti-15V-3Cr-3Al-3Sn | | STA | 1,230 | 1,110 | 10 |

A summary for each irradiation phase is given in Table II. Beam exposure was 55 days in total, and the integrated pot reached $4.6\times10^{21}$, which is about two times larger than the pot expected for 1.3MW operation per year at the J-PARC neutrino facility (cf. Table I). Peak damage for DS-Ti1, DS-Ti2 and US-Ti were 0.25, 0.96 and 1.5 dpa, respectively, based on the estimation by NRT model[15].

*2.2 Classification of Titanium alloys and BLIP Specimens*

Properties of an alloy material, particularly its mechanical properties, such as elastic and plastic deformation behavior, are dependent on its crystalline microstructure. At ambient temperature and pressure, pure titanium is in HCP (α-phase), and undergoes an allotropic transformation to BCC (β-phase) at 882°C. The addition of Al/Zr/Sn… (α-stabilizer) raises the temperature of β-transus, whereas the addition of V/Nb/Mo… (β-stabilizer) lowers it, as schematically shown in the pseudo-binary β-isomorphous phase diagram, Fig. 4. By combining these elements, titanium alloys can be designed to exhibit a variety of phase structures, which are ordinarily classified into three groups: single-α phase, dual α+β phase, and meta-stable β phase alloys. The α+β phase and metastable β phase alloys are heat-treatable, which is referred to as Solution Treatment and Aging (STA): maintain the alloy at a temperature higher than the β transus to dissolve the alloying elements (Solution Treatment), then quench it to keep it in a metastable supersaturated solid solution state. After the ST process, aging at an elevated temperature for several hours generates a fine (less than 10 nm) scale precipitation in β-phase grains, i.e., α phase for aging above ~500°C, or ω-phase for aging between 400~500 °C. Whilst the fine scale precipitation makes the alloy stronger ("precipitation hardening"), there is evidence to suggest that the ω precipitation can lead to embrittlement ("ω-embrittlement"). It is to be noted that, so far radiation damage effects on titanium alloys have only been studied on standard grades, such as mill-annealed (A) Ti-6Al-4V and a few α-phase alloys such as Ti-5Al-2.5Sn[11][16]. It is of interest how this wide variety of phase compositions may affect the radiation damage behavior on their mechanical properties. For example, after irradiation to 0.3 dpa, significant ductility remains for α-alloy Ti-5Al-2.5Sn in contrast to the α+β alloy Ti-6Al-4V, where no uniform elongation remains after the same level of radiation damage[11][16]. It has been suggested that nano-scale precipitates generated in the metastable β phase alloy such as Ti-15V-3Cr-3Al-3Sn[17], and/or rich grain boundaries of ultrafine-grained microstructure introduced by special heat treatments[18], may act as "sink-sites" for point defects caused by radiation damage. Specimens of these potential radiation damage tolerant candidates were introduced into one of the BLIP irradiation capsules, DS-Ti2.

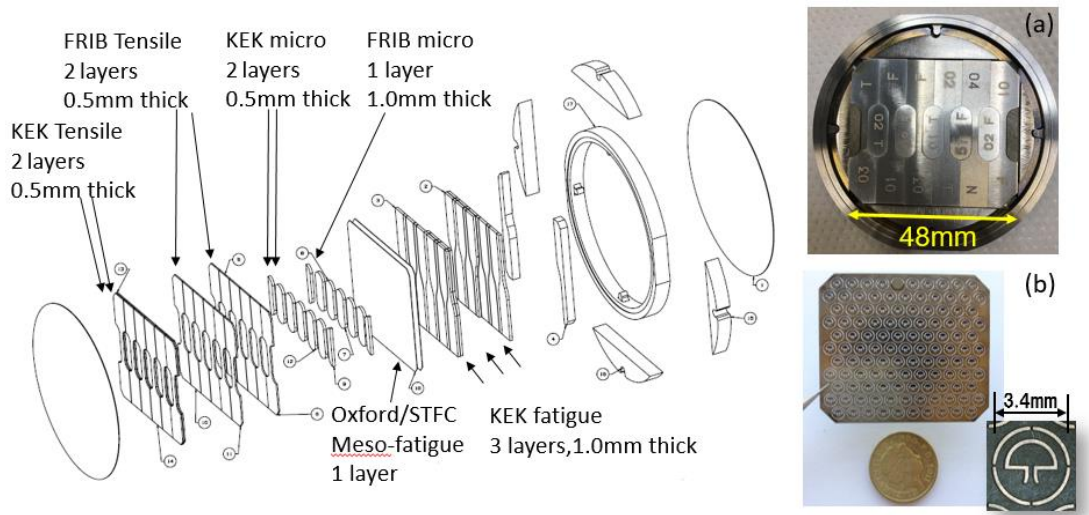

**Fig. 5.** Assembly of the specimens in the US-Ti capsule. (a) Picture for the upstream tensile + microstructural specimen layer. (b) Meso-scale fatigue testing foil: numerous small mushroom-shaped cantilevers are laser-machined in 0.1~0.2 mm-thick foil, to be utilized for ultrasonic HCF testing.

In the BLIP capsules, a variety of specimens were arranged in multiple layers. As an example, Fig. 5 shows an expanded view of specimens in the US-Ti capsule. It contains four 0.5 mm-thick layers of 6 tensile specimens upstream, where each has a 42 mm-long dog-bone shape with shoulder loading, with a gauge cross-section 0.5 mm-thick and 1.75 mm-wide. The five gaps between the neighboring tensile specimens are filled with oval-shaped specimens 0.5 mm and 1 mm thick for microstructural characterization, such as SEM, EBSD, TEM, AFM etc (Fig. 5(a)), to be reported as separate works. The capsule also contains three 1 mm-thick layers of 10 bend fatigue samples downstream, which are to be tested at Fermilab by a newly-developed compact fatigue testing machine with remote handling capability[19]. Between the upstream tensile and the downstream fatigue specimen layers, a 0.25 mm-thick Ti-6Al-4V Grade-23 foil, with numerous small laser-machined cantilevers was installed for ultrasonic fatigue testing (Fig. 5(b)). This is an innovative technique at Oxford University[20] and now being developed at Culham Laboratory specifically to test these irradiated fatigue specimens in order to provide the first HCF data for irradiated titanium alloys.

*2.3 Status of Post-Irradiation Examination (PIE) at PNNL*

After the phase-1 irradiation in 2017, the DS-Ti1 capsule experienced a total of $7.30 \times 10^{20}$ pot, and the accumulated radiation damage reached 0.25 dpa at beam profile center, which is comparable to the existing irradiation data. After a few months of cool-down, it was shipped to PNNL in the end of CY2017 for PIE studies. The capsule was opened in a hot-cell by using a specially-developed remote capsule opener, and a series of remote tensile tests were performed.

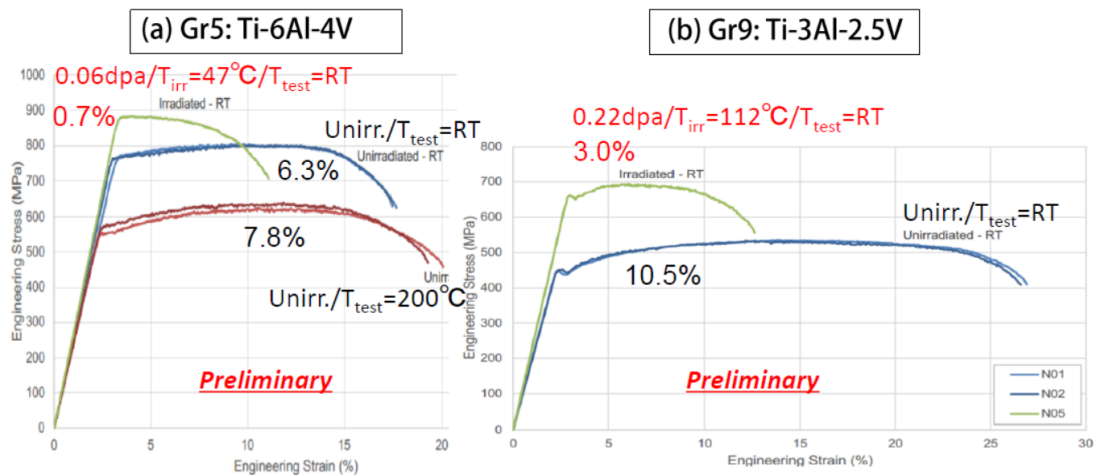

**Fig. 6.** (a) Stress-strain curves of ASTM Grade-5 Ti-6Al-4V, tests at Room Temperature (RT) and at 200°C, for irradiated/un-irradiated specimens. (b) Stress-strain curves of ASTM Grade-9 Ti-3Al-2.5V, tested at RT for irradiated and un-irradiated specimens. The percentage numbers stand for the uniform elongation.

As an example, the stress-strain curves for the α+β alloy Ti-6Al-4V ASTM Grade-5, both irradiated and unirradiated control specimens, are plotted together in Fig. 6(a). Unirradiated specimens exhibit modest work hardening capability at both test temperatures (RT and at 200°C) and the failure mode is ductile. However, a test at RT for an irradiated specimen, located at the edge of the capsule and thus only minimally irradiated (0.06 dpa at $T_{irr}$=47°C), indicates an increase in the yield stress (nearly identical to the ultimate tensile strength) with a remarkably large reduction in uniform elongation (6.3% to 0.7%). Thus this material has lost its ability to work harden to any significant extent. On the other hand, the stress-strain curves for the near-α α+β alloy Ti-3Al-2.5V ASTM Grade-9 specimens tested at RT are plotted in Fig. 6(b). For the control measurements, the yield and tensile strength are lower than those for Ti-6Al-4V, while the total elongation is much larger. For a specimen irradiated to 0.22 dpa at $T_{irr}$=112°C, the increase in yield stress of more than 200 MPa is about a factor of 2 larger than that for the Grade-5. Although a large decrease of both uniform and total elongation is observed, modest work hardening capability still remains after irradiation, with a uniform elongation of 3.0%. To clarify what microstructural difference causes the larger decrease in ductility for Ti-6Al-4V than that for Ti-3Al-2.5V, investigations are in progress at PNNL on the microstructural specimens. A comprehensive report for the DS-Ti1 tensile and microstructural PIE is under preparation [21].

## 3. Thermal Shock Study at CERN HiRadMat Facility

The RaDIATE collaboration also contributed to thermal shock studies on target and beam window materials at CERN's HiRadMat facility: HiRadMat[22], High-Radiation to Materials, is a user facility designed to perform single to several beam impingements on materials and accelerator component assemblies to evaluate the effect of high-intensity pulsed beams in a controlled environment. Using an existing fast extraction channel to the LHC, a high-intensity proton beam from CERN SPS with 440 GeV beam energy is delivered to the facility, with a maximum pulse intensity of $3.5 \times 10^{13}$ in a pulse length of 7.2 µs, being equivalent to 3.4 MJ total beam energy. The beam is collimated at the target with a Gaussian spot size of σ tunable between 0.1 and 2 mm. This highly-collimated beam shot generates an extremely high thermal stress in materials, which can potentially cause fracture even in graphite and beryllium. The HRMT43-BeGrid2 experiment[23] is a follow-up of a past experiment HRMT24-BeGrid[24], aiming to expose beryllium, a typical beam window material frequently used at Fermilab and other facilities, to even higher beam intensities than previously achieved. The main purpose of BeGrid2 was to expose material specimens previously irradiated at BLIP to compare the thermal shock response with that for non-irradiated specimens. Figure 7 shows the experimental setup of BeGrid2. Irradiated specimens, including beryllium, graphite, silicon, titanium, SiC-coated graphite, and glassy carbon, were assembled into three irradiation array boxes in PNNL's hot-cell, where the box#3 contained irradiated Titanium alloy Grade-5 and Grade-9, together with SiC-coated graphite and Silicon specimens. These boxes were then shipped to CERN, and were integrated into the experimental setup at HiRadMat facility tunnel. The beam exposures were successfully conducted on October 1 and 2, 2018. After cool-down and shipment to CCFE (UK), the array#3 specimens will be shipped back to PNNL to perform PIE.

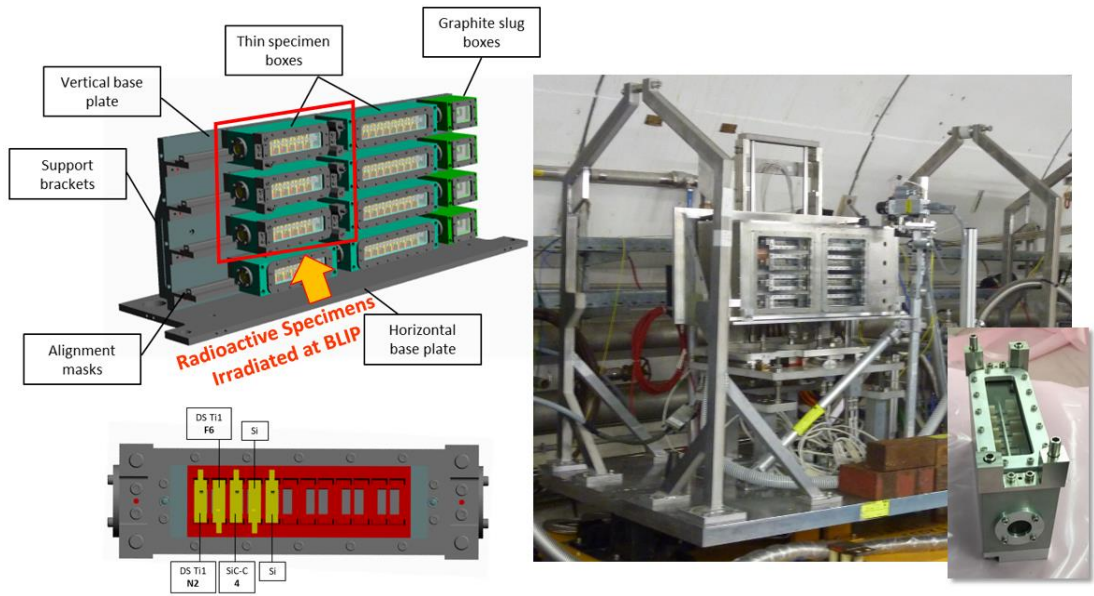

**Fig. 7.** Setup for the BeGrid2 experiment. 8 thin specimen boxes, including 3 with radioactive specimens previously irradiated at BLIP, are assembled into a movable assembly. Each specimen box and the entire assembly are hermetically sealed, to fully contain specimens and to avoid contamination release. The array box #3 contains irradiated Titanium alloy Gr-5 and Gr-9.

## 4. Summary and Prospects

- During 2017 to 2018, the RaDIATE collaboration conducted a high-intensity proton beam irradiation campaign at the BNL-BLIP facility. The accumulated pot and radiation damage are comparable to the annual level for future MW-class facility operation.
- Post-Irradiation Examination for titanium alloys is in progress on the 1st capsule irradiated in 2017 (DS-Ti1). Tensile tests on Ti-3Al-2.5V Grade-9 with less V/Al alloy elements (less β phase) exhibit better ductility than Ti-6Al-4V Grade-5 or ELI Grade-23. To reveal the cause of this difference in tensile behavior, intensive microstructural investigations are now in progress at PNNL.
- Thermal shock tests on target materials irradiated at BLIP was also performed at CERN HiRadMat facility. It is to be emphasized that testing of highly irradiated samples subjected to thermal shock from such a high intensity beam has never been conducted before.
- The first High-Cycle Fatigue data, which is critical for beam window applications, are to be obtained by a bend-fatigue test at Fermilab and meso-scale ultrasonic tests at Culham laboratory.
- The second capsule irradiated in 2018 (DS-Ti2) at up to 1 dpa contained a variety of grades, to cover typical titanium alloy grades and possible radiation-resistant candidates. Systematic studies of radiation damage effects on titanium alloys are intended to enable us to predict realistic lifetimes of existing beam windows made of Ti-6Al-4V, and to identify grades and heat treatments that may extend the lifetime of such components.

**Acknowledgment**

This work was financially supported by the U.S.-Japan Science and Technology Cooperation Program in High Energy Physics. This manuscript has been authored by Fermi Research Alliance, LLC under Contract No. DE-AC02-07CH11359 with the U.S. Department of Energy, Office of Science, Office of High Energy Physics.